\documentclass[10pt,twocolumn,aps,pra,showpacs]{revtex4}

\usepackage{bbm}

\begin{document}

\title{General framework for quantum search algorithms}         
\author{Avatar Tulsi\\
        {\small Department of Physics, Indian Institute of Science, Bangalore-560012, India}}  

\email{tulsi9@gmail.com}

\begin{abstract}
Grover's quantum search algorithm drives a quantum computer from a prepared initial state to a desired final state by using selective transformations of these states. Here, we analyze a framework when one of the selective trasformations is replaced by a more general unitary transformation. Our framework encapsulates several previous generalizations of the Grover's algorithm. We show that the general quantum search algorithm can be improved by controlling the transformations through an ancilla qubit. As a special case of this improvement, we get a faster quantum algorithm for the two-dimensional spatial search.  
\end{abstract}


\maketitle

\section{Introduction}

The search problem is to find a desired item satisfying certain properties out of a given database of $N$ items. Consider a quantum system with an $N$-dimensional Hilbert space $\mathcal{H}_{N}$, whose basis states $|j\rangle$, $j \in \{0,1,\ldots,N-1\}$, encode the $N$ items of the database. The target state $|t\rangle$ corresponds to the desired item, while all other basis states are non-target states. In general, the system can be in any normalized superposition of the basis states. We initialize the system in the source state $|s\rangle$ and assume $\alpha  = |\langle t|s\rangle|$ to be non-zero. Direct measurement of $|s\rangle$ yields the target state $|t\rangle$ with a probability $\alpha^{2}$.  So on average, $1/\alpha^{2}$ preparations of $|s\rangle$ and subsequent measurements will yield $|t\rangle$ with high probability. 

Grover's algorithm, or more generally quantum amplitude amplification~\cite{grover,qaa1,qaa2}, makes the search faster, by performing measurement only after transforming $|s\rangle$ to $|t\rangle$ using $\pi/4\alpha$ iterations of the unitary operator $\mathcal{G} = -I_{s}I_{t}$, where $I_{\psi} = \mathbbm{1}_{N}-2|\psi\rangle\langle \psi|$ is the selective phase inversion of the state $|\psi\rangle$. That is $O(1/|\langle t|s\rangle|)$ times faster than the measurement scheme. Generally, $|s\rangle$ is prepared by applying a unitary operator $U$ on a particular basis state, say $|0\rangle$. Then $|s\rangle = U|0\rangle$, and $|t\rangle$ is obtained by iterating the operator $\mathcal{G} = -UI_{0}U^{\dagger}I_{t}$ on $U|0\rangle$. Grover's algorithm has been proved to be optimal, i.e. no other algorithm can get to $|t\rangle$ faster than it does~\cite{optimal}. 

To search a database using Grover's algorithm, we choose $|s\rangle$ to be the equal superposition of all the basis states, i.e. $|s\rangle = \sum_{j}|j\rangle/\sqrt{N}$. Such a state is generated by applying the Walsh-Hadamard transformation $W$ on the basis state $|0\rangle$, i.e. $|s\rangle = W|0\rangle$. Then Grover's algorithm iterates the operator $\mathcal{G} = -I_{s}I_{t} = -WI_{0}WI_{t}$ on $|s\rangle$ to reach $|t\rangle$. If $|t\rangle$ is a unique basis state, $\alpha = 1/\sqrt{N}$, and $|t\rangle$ is reached using $\lfloor \pi\sqrt{N}/4 \rfloor$ queries.   
 
Grover's algorithm has been generalized in several ways. One generalization is to replace $\mathcal{G}$ by $\mathcal{G}_{\varphi, \phi} = e^{-i\varphi}I_{s}^{\varphi}I_{t}^{\phi}$, where $I_{\psi}^{\omega} = \mathbbm{1}_{N}-\left(1-e^{\imath\omega}\right)|\psi\rangle\langle \psi|$ is the selective phase rotation of $|\psi\rangle$ by angle $\omega$ (clearly $I_{\psi}^{\pi} \equiv I_{\psi}$). Then $|t\rangle$ is reached by applying $O(\csc \frac{|\phi|}{2}/\alpha)$ iterations of $\mathcal{G}_{\varphi,\phi}$ to $|s\rangle$, provided the phase-matching condition $\varphi-\phi = O(\alpha)$ is satisfied~\cite{phase1,phase2}. Kato's algorithm~\cite{kato} replaces $I_{s}$ by an operator $K_{s}$ that, unlike $I_{s}$, consists of only single qubit operators and hence is easier to implement physically. Ambainis' algorithm~\cite{ambainis} replaces $I_{s}$ by a real operator $R_{s}$, having $|s\rangle$ as its unique eigenstate with eigenvalue $1$. It has important applications in fast quantum walk algorithms for element distinctness~\cite{ambainis} and spatial search~\cite{spatiald,spatialc,aaronson}. 

Here we study further generalizations of Grover's algorithm. Section II presents and analyses a general quantum search algorithm that iterates the search operator $\mathcal{S} = D_{s}I_{t}^{\phi}$ on $|s\rangle$ to reach $|t\rangle$, where $D_{s}$ can be almost any unitary operator with $|s\rangle$ as its eigenstate. We find the conditions that $D_{s}$ must satisfy to yield a successful quantum search algorithm, i.e. reach $|t\rangle$ from $|s\rangle$ using $O(1/\alpha)$ iterations of $\mathcal{S}$. Section III illustrates the general algorithm by explaining Grover's algorithm, phase-matching condition, Kato's algorithm, and spatial search as its special cases. Section IV presents a controlled algorithm where the $\{D_{s},I_{t}^{\phi}\}$ operators are controlled through an ancilla qubit to speed up the search. It provides an important application for the two dimensional spatial search problem, solving it in $O(\sqrt{N\ln N})$ time steps~\cite{faster}. Earlier algorithms solved this problem in $O(\sqrt{N}\ln N)$ time steps, and it was an open question to design a faster algorithm. Another application is to find an eigenstate of a given unitary operator, corresponding to a known eigenvalue. These results are easily extended, by time reversal symmetry, to problems with interchanged roles $s\leftrightarrow t$.

\section{GENERAL QUANTUM SEARCH ALGORITHM}

\subsection{General search operator}

In this section, we analyse a general quantum search algorithm that iterates the search operator $\mathcal{S} = D_{s}I_{t}^{\phi}$ on $|s\rangle$ to take it close to $|t\rangle$. Here $D_{s}$ is a unitary operator with the initial state $|s\rangle$ as its eigenstate. We want to derive conditions on the eigenspectrum of $D_{s}$ that allow an efficient construction of the search algorithm. Since a global phase is irrelevant in quantum dynamics, we choose the convention $D_{s}|s\rangle = |s\rangle$. In case $D_{s}$ has an $M$-dimensional degenerate eigenspace with eigenvalue $1$, orthonormally spanned by $|s_{m}\rangle$ with $m \in \{1,2,\ldots,M\}$, we choose the initial state $|s\rangle$ to be
\begin{equation}
|s\rangle = \frac{1}{\alpha}\sum_{m}\langle s_{m}|t\rangle|s_{m}\rangle\ ,\ \ \alpha^{2} = \sum_{m}|\langle s_{m}|t\rangle|^{2}\ . \label{initialstatedefine}
\end{equation}
Note that $\alpha = |\langle t|s\rangle|$. We consider $\alpha \ll 1$ so that the quantum search algorithm provides a significant speedup, $O(1/\alpha)$, over classical search algorithms. 

Let the normalized eigenstates $|\ell\rangle$ of $D_{s}$ satisfy
\begin{eqnarray}
&D_{s}|\ell\rangle = e^{\imath\theta_{\ell}}|\ell\rangle,\ \ D_{s}^{\dagger}|\ell\rangle = e^{-\imath \theta_{\ell}}|\ell\rangle,&  \nonumber \\ 
&\theta_{\ell=s_{m}} = 0,\ \ |\theta_{\ell\neq s_{m}}| \geq \theta_{\rm min} > 0,\ \ \theta_{\ell} \in [-\pi,\pi].& \label{Dsdefine}
\end{eqnarray}
Here the lower bound $\theta_{\rm min}$ characterizes the gap in the eigenspectrum of $D_{s}$.
 
To analyse the iteration of $\mathcal{S}$ on $|s\rangle$, we need to find its eigenvalues and eigenvectors. Let $|\lambda_{k}\rangle$ be the normalized eigenvectors of $\mathcal{S}$ with eigenvalues $e^{\imath \lambda_{k}}$. Now, the definition $I_{t}^{\phi} = \mathbbm{1}_{N} - (1-e^{\imath \phi})|t\rangle \langle t|$ implies that $I_{t}^{\phi}|\lambda_{k}\rangle = |\lambda_{k}\rangle - (1-e^{\imath\phi})\langle t|\lambda_{k}\rangle |t\rangle$. Left multiplication by $\langle \ell |D_{s} = e^{\imath\theta_{\ell}}\langle \ell|$ gives
\begin{eqnarray}
\langle \ell|D_{s}I_{t}^{\phi}|\lambda_{k}\rangle&=&e^{\imath\lambda_{k}}\langle \ell|\lambda_{k}\rangle \nonumber \\
&=&e^{\imath\theta_{\ell}}\left[\langle \ell|\lambda_{k}\rangle - (1-e^{\imath\phi})\langle t|\lambda_{k}\rangle \langle \ell|t\rangle\right]. \nonumber 
\end{eqnarray}
Thus 
\begin{equation}
\langle \ell|\lambda_{k}\rangle = \langle t|\lambda_{k}\rangle\langle \ell|t\rangle  \frac{1-e^{\imath\phi}}{1-e^{\imath(\lambda_{k}-\theta_{\ell})}}\ . \label{lmcoeff}
\end{equation}
Using $\langle t|\lambda_{k}\rangle = \sum_{\ell}\langle t|\ell\rangle\langle \ell|\lambda_{k}\rangle$, we get
\begin{equation}
\langle t|\lambda_{k}\rangle = \langle t|\lambda_{k}\rangle (1-e^{\imath \phi})\sum_{\ell}\frac{|\langle \ell|t\rangle|^{2}}{1-e^{\imath(\lambda_{k}-\theta_{\ell})}}\ , \label{tlambdapmnew}
\end{equation}
which leads to
\begin{displaymath}
\sum_{\ell}\frac{|\langle \ell|t\rangle|^{2}}{1-e^{\imath(\lambda_{k}-\theta_{\ell})}} = \frac{1}{1-e^{\imath\phi}}
\end{displaymath}
or
\begin{equation}
\sum_{\ell}|\langle \ell|t\rangle|^{2}\cot\frac{\lambda_{k}-\theta_{\ell}}{2} = \cot\frac{\phi}{2}\ . \label{eigencondition}
\end{equation}
The last result follows from $(1-e^{\imath x})^{-1} = \frac{1}{2}+\frac{\imath}{2} \cot\frac{x}{2}$ and the normalization condition $\sum_{\ell}|\langle \ell|t\rangle|^{2} = 1$. It is the essential condition for $e^{\imath \lambda_{k}}$ to be an eigenvalue of $\mathcal{S}$. 

Since the R.H.S. of (\ref{eigencondition}) is a constant, and $\cot x$ varies monotonically with $x$ except for the jump from $-\infty$ to $+\infty$ when $x$ crosses zero, there is a unique solution $\lambda_{k}$ between each pair of consecutive $\theta_{\ell}$'s. Then with $\theta_{\ell =s_{m}} = 0$, there can be at most two solutions $\lambda_{k}$ in the interval $[-\theta_{\rm min},\theta_{\rm min}]$. It is also clear that one can focus on the solutions $\lambda_{k}$ on either side of any particular $\theta_{\ell}$, by a suitable choice of the inconsequential global phase, i.e. by considering the search operator $e^{-\imath \theta_{\ell}}D_{s}I_{t}^{\phi}$. 

\subsection{Eigenvalues of the general search operator}

To explicitly solve the secular equation (\ref{eigencondition}), we assume that the smallest eigenphases of $\mathcal{S}$ obey
\begin{equation}
{\rm Assumption}:\ \ \ |\lambda_{k}| \ll \theta_{\rm min}\ \  \Longrightarrow\ \  |\lambda_{k}| \ll 1. \label{assumption}
\end{equation}
It makes the $\ell =s_{m}$ contribution the most significant one in the sum on L.H.S. of (\ref{eigencondition}). Now we use the Laurent series expansions, ignoring $O(\lambda_{k}^{2})$ terms, $\cot\frac{\lambda_{k}}{2} = \frac{2}{\lambda_{k}}-\frac{\lambda_{k}}{6}$ and
\begin{displaymath}
\cot\frac{\lambda_{k}-\theta_{\ell}}{2} = -\cot \frac{\theta_{\ell}}{2} - \frac{\lambda_{k}}{2}\left(1+\cot^{2} \frac{\theta_{\ell}}{2}\right).
\end{displaymath}
Along with $\theta_{\ell =s_{m}} = 0$ and $\sum_{m}|\langle s_{m}|t\rangle|^{2} =\alpha^{2}$, (\ref{eigencondition}) then reduces to a quadratic equation
\begin{equation}
\frac{2\alpha^{2}}{\lambda_{k}} - A - \frac{B^{2}}{2}\lambda_{k} =  0. \label{quadratic} 
\end{equation}
Its two solutions are the expected eigenphases on either side of $\theta_{\ell =s_{m}}=0$ and satisfying (\ref{assumption}). The coefficients $A,B$ are, up to $O(\alpha^{2},\lambda_{k}^{2})$,
\begin{equation}
A = \cot\frac{\phi}{2}+\Lambda_{1}\ ,\ \ B^{2} = 1+\Lambda_{2}\ , \label{ABdefine}
\end{equation}
where
\begin{equation}
\Lambda_{p} = \sum_{\ell\neq s_{m}}|\langle \ell|t\rangle|^{2}\cot^{p} \frac{\theta_{\ell}}{2}\ ,\ p\in\{1,2\}, \label{Omega}
\end{equation}
is the $p^{th}$ moment of $\cot\frac{\theta_{\ell}}{2}$ with respect to the distribution $|\langle \ell |t\rangle|^{2}$ over all $\ell \neq s_{m}$. While $\Lambda_{1}$ can have any sign, $\Lambda_{2}$ is always positive. Since $\theta_l \in [-\pi,\pi]$, we have $|\cot \frac{\theta_{\ell}}{2}| \leq \frac{2}{|\theta_{\ell}|} \leq \frac{2}{\theta_{\rm min}}$ for $\ell \neq s_{m}$. Then $\sum_{\ell \neq s_{m}} |\langle\ell|t\rangle|^{2} = 1-\alpha^{2} \leq 1$ gives the bounds
\begin{equation}
|\Lambda_{p}| \leq (2/\theta_{\rm min})^{p},\ \ B^{2}\ \in\ [1,1+(4/\theta_{\rm min}^{2})]\ .\label{Bbound}
\end{equation}
Moreover, putting $x_{\ell} = |\langle \ell|t\rangle|$ and $y_{\ell} = |\langle \ell|t\rangle|\cot\frac{\theta_{\ell}}{2}$ in the Cauchy-Schwartz inequality, $(\sum x_{\ell}y_{\ell})^{2}\ \leq\ (\sum x_{\ell}^{2})(\sum y_{\ell}^{2})$, gives
\begin{equation}
\Lambda_{1}^{2} \leq (1-\alpha^{2})\Lambda_{2} < \Lambda_{2}\ .
\end{equation}

Let the two roots of (\ref{quadratic}) be $\lambda_{\pm}$. Their product is $\lambda_{+}\lambda_{-} = -4\alpha^{2}/B^{2}$ so we can write
\begin{equation}
\lambda_{\pm} = \pm\frac{2\alpha}{B}(\tan \eta)^{\pm 1}\ ,\ \ \eta \in \left[0,\frac{\pi}{2}\right].\label{solutions}
\end{equation}
The sum of the two roots determines the angle $\eta$. We have $\lambda_{+}+\lambda_{-} = -2A/B^{2} = (2\alpha/B)(\tan \eta - \cot \eta) = -(4\alpha/B)\cot 2\eta$. Thus
\begin{equation}
\cot 2\eta = \frac{A}{2\alpha B}\ .  \label{etadefine} 
\end{equation}
For all $\eta$, we have $|\tan\eta-\cot\eta| \leq \max_{\pm}(\tan \eta)^{\pm 1} \leq 1+|\tan\eta-\cot\eta|$.  Together with $|\tan\eta-\cot\eta| = |2\cot 2\eta|$, we then obtain
\begin{equation}
2|A|/B^{2} \leq {\textstyle \max_{\pm}}|\lambda_{\pm}| \leq (2\alpha/B)+(2|A|/B^{2}). \label{lambdabound}
\end{equation}
Thus the assumption (\ref{assumption}) for the validity of our analysis necessarily holds, when
\begin{equation}
2\alpha B + 2|A| \ \ll\  \theta_{\rm min}B^{2} \ \in\  [\theta_{\rm min}\ ,\ \theta_{\rm min}+4\theta_{\rm min}^{-1}]\ , \label{validitycondition2}
\end{equation}
making use of (\ref{Bbound}). With $\alpha B \geq 0$ and $\theta_{\rm min} \leq \pi$, we certainly need $|A| \ll \frac{\pi^{2}+4}{2\theta_{\rm min}}$.

In general, the validity condition (\ref{validitycondition2}) involves parameters of the database, $D_{s}$ and $I_{t}^{\phi}$. We will look at several special cases in Sections III and IV.

\subsection{Eigenvectors of the general search operator}

We calculate the eigenstates $|\lambda_{\pm}\rangle$, again dropping $O(\alpha^{2},\lambda_{\pm}^{2})$ terms from the analysis throughout. We choose $|\lambda_{\pm}\rangle$ such that $\langle t|\lambda_{\pm}\rangle$ are real and positive. Then the normalization condition $\sum_{\ell}|\langle \ell|\lambda_{\pm}\rangle|^{2} =1$ in (\ref{lmcoeff}) gives
\begin{equation}
\frac{\csc^{2}\frac{\phi}{2}}{\langle t|\lambda_{\pm}\rangle^{2}} = \sum_{\ell}|\langle \ell|t\rangle|^{2}\csc^{2}\frac{\lambda_{\pm}-\theta_{\ell}}{2}\ . \label{normalizing}
\end{equation}
In the sum over $\ell$, the $\ell =s_{m}$ terms contribute as per (\ref{solutions}),
\begin{equation}
\alpha^{2}\csc^{2}{\textstyle \frac{\lambda_{\pm}}{2}} = 4\alpha^{2}/\lambda_{\pm}^{2} = B^{2}(\tan \eta)^{\mp 2}.
\end{equation} 
To find the sum of $\ell \neq s_{m}$ terms, we use the assumption (\ref{assumption}) to get the expansion $\csc^{2}{\textstyle \frac{\lambda_{\pm}-\theta_{\ell}}{2}} = \csc^{2}{\textstyle \frac{\theta_{\ell}}{2}}\left(1+\lambda_{\pm}\cot{\textstyle \frac{\theta_{\ell}}{2}}\right)$. Then 
\begin{equation}
\sum_{\ell \neq s_{m}}|\langle \ell|t\rangle|^{2}\csc^{2}{\textstyle \frac{\lambda_{\pm}-\theta_{\ell}}{2}} = B^{2}\left[1+O\left({\textstyle \frac{\lambda_{\pm}}{\theta_{\rm min}}}\right)\right].
\end{equation}
Here we have used the inequality $|\cot\frac{\theta_{\ell \neq s_{m}}}{2}| \leq \frac{2}{\theta_{\rm min}}$ and the expression $B^{2} = \sum_{\ell \neq s_{m}}|\langle \ell|t\rangle|^{2}\csc^{2}{\textstyle \frac{\theta_{\ell}}{2}}$ which follows from (\ref{ABdefine}) and (\ref{Omega}) with $\csc^{2}\frac{\theta_{\ell}}{2} = 1+\cot^{2}\frac{\theta_{\ell}}{2}$. Ignoring $O(\lambda_{\pm}/\theta_{\rm min})$ contribution in accordance with (\ref{assumption}), (\ref{normalizing}) then gives
\begin{displaymath}
\langle t|\lambda_{\pm}\rangle^{2} = \frac{\csc^{2}\frac{\phi}{2}}{B^{2}[1+(\tan \eta)^{\mp 2}]}
\end{displaymath}
or
\begin{equation}
\langle t|\lambda_{\pm}\rangle = \frac{|\csc\frac{\phi}{2}|}{B\sqrt{1+(\tan \eta)^{\mp 2}}}\ .\label{tlambdapm1}
\end{equation}
Thus
\begin{equation}
|t\rangle = \frac{|w\rangle}{B|\sin\frac{\phi}{2}|} +|\lambda_{\perp}\rangle\ ,\ |w\rangle = \sin\eta |\lambda_{+}\rangle + \cos \eta |\lambda_{-}\rangle, \label{texpand}
\end{equation} 
where $|w\rangle$ is the normalized projection of $|t\rangle$ on the $|\lambda_{\pm}\rangle$-subspace, and $|\lambda_{\perp}\rangle$ is an unnormalized component orthogonal to this subspace. 

Note that $\|\lambda_{\perp}\|^{2} = 1-B^{-2}\csc^{2}\frac{\phi}{2} \geq 0$ requires $B^{-2}\csc^{2}\frac{\phi}{2} \leq 1$. With $B^{2} = 1+\Lambda_{2}$ and $\csc^{2}x = 1+\cot^{2}x$, it is equivalent to having $\cot^{2}\frac{\phi}{2} \leq \Lambda_{2}$. That trivially holds for $\phi=\pi$. In general, $\phi$ and $\Lambda_{2}$ are mutually independent, arising from $I_{t}^{\phi}$ and $D_{s}$ respectively, and the constraint has to follow from the assumption $\lambda_{\pm} \ll \theta_{\rm min}$. In our analysis, both $|A|$ and $|\Lambda_{1}|$ are bounded by constant multiples of $\theta_{\rm min}^{-1}$. Therefore,
\begin{eqnarray}
\cot^{2}{\textstyle \frac{\phi}{2}} &=& \Lambda_{1}^{2} + A(A-2\Lambda_{1}) \nonumber \\ 
&=& \Lambda_{1}^2 + O\left(\frac{|A|}{\theta_{\rm min}}\right) \ \leq\ \Lambda_{2} + O\left(\frac{B^{2}|\lambda_{\pm}|}{\theta_{\rm min}}\right) \nonumber, \label{indinequal}
\end{eqnarray} 
where the last inequality follows from $\Lambda_{1}^2 < \Lambda_2$ and (\ref{lambdabound}). Thus we indeed have $\cot^{2}\frac{\phi}{2} \leq \Lambda_{2}[1+O(\lambda_{\pm}/\theta_{\rm min})]$, consistent with the $O(\lambda_{\pm}/\theta_{\rm min})$ accuracy of our analysis.

Putting (\ref{tlambdapm1}) in (\ref{lmcoeff}), we obtain the desired eigenvectors, $|\lambda_{\pm}\rangle = \sum_{\ell}\langle \ell|\lambda_{\pm}\rangle|\ell\rangle$, or
\begin{equation}
|\lambda_{\pm}\rangle = \frac{e^{\imath (\phi-\lambda_{\pm})/2}}{B\sqrt{1+(\tan \eta)^{\mp 2}}}\ \sum_{\ell}\langle \ell|t\rangle \csc\frac{\lambda_{\pm}-\theta_{\ell}}{2}e^{\imath \theta_{\ell}/2}\ |\ell\rangle\ . \label{eigenstateexpression}
\end{equation}
To find the initial state $|s\rangle$ in terms of $|\lambda_{\pm}\rangle$, we calculate the projections $\langle s|\lambda_{\pm}\rangle$. Using (\ref{initialstatedefine}) and $\theta_{\ell = s_{m}} = 0$, together with (\ref{solutions}) and (\ref{eigenstateexpression}), we get
\begin{eqnarray}
\langle s|\lambda_{\pm}\rangle &=& \frac{e^{\imath (\phi-\lambda_{\pm})/2}}{B\sqrt{1+(\tan \eta)^{\mp 2}}}\frac{1}{\alpha}\sum_{m}|\langle s_{m}|t\rangle|^{2}\csc\frac{\lambda_{\pm}}{2} \nonumber \\
&=&  \pm \frac{e^{\imath (\phi-\lambda_{\pm})/2}}{\sqrt{1+(\tan \eta)^{\pm 2}}}\ .
\end{eqnarray}
We observe that $\sum_{\pm}|\langle s|\lambda_{\pm}\rangle|^{2} =1$, and so we do not have to worry about component of $|s\rangle$ orthogonal to the $|\lambda_{\pm}\rangle$-subspace. Explicitly, we have
\begin{equation}
|s\rangle = e^{-\imath \phi/2}[e^{\imath \lambda_{+}/2}\cos \eta |\lambda_{+}\rangle - e^{\imath \lambda_{-}/2}\sin \eta |\lambda_{-}\rangle], \label{slambdapmexpansion}
\end{equation}
which yields
\begin{equation}
\mathcal{S}^{q}|s\rangle =  e^{-\imath \phi/2}[e^{\imath q'\lambda_{+}}\cos \eta |\lambda_{+}\rangle - e^{\imath q'\lambda_{-}}\sin \eta |\lambda_{-}\rangle],
 \label{stateexpand}
\end{equation}
where $q'  = q+\frac{1}{2}$.

The results of this Section hold up to $O(\alpha^{2},\lambda_{\pm}/\theta_{\rm min})$, and to that extent the $|\lambda_{\pm}\rangle$-subspace suffices for our analysis. We note that the eigenvalues are determined to higher accuracy than the eigenvectors, as is common in unitary quantum mechanics. 

\subsection{Algorithm's performance}

The success probability of the algorithm, i.e. obtaining $|t\rangle$ upon measuring $\mathcal{S}^{q}|s\rangle$, is $P_{t}(q)=|\langle t|\mathcal{S}^{q}|s\rangle|^{2}$. From (\ref{texpand}) and (\ref{stateexpand}), 
\begin{equation}
|\langle t|\mathcal{S}^{q}|s\rangle| = \frac{\sin 2\eta |e^{\imath q'\lambda_{+}}-e^{\imath q'\lambda_{-}}|}{2B|\sin\frac{\phi}{2}|} \ .
\end{equation}
With $\Delta\lambda \equiv \lambda_{+}-\lambda_{-} = 4\alpha/(B\sin 2\eta)$, we get
\begin{equation}
|\langle t|\mathcal{S}^{q}|s\rangle|  = \frac{\sin 2\eta}{B|\sin\frac{\phi}{2}|}\sin\left(\frac{2q'\alpha}{B\sin 2\eta}\right). \label{tSqs}
\end{equation}
As the maximum value of $\sin x$ is $1$ for $x = \pi/2$, the maximum success probability $P_{\rm m}$, and the number of iterations $q_{\rm m}$ needed to achieve it, are
\begin{displaymath}
|u\rangle =\mathcal{S}^{q_{\rm m}}|s\rangle,\ \ \ P_{\rm m} = |\langle t|u\rangle|^{2} = \frac{\sin^{2}2\eta}{B^{2}\sin^{2}\frac{\phi}{2}}\ ,
\end{displaymath}
\begin{equation} 
q_{\rm m} =\left\lfloor\frac{\pi}{\lambda_{+}-\lambda_{-}}\right\rfloor =  \left\lfloor\frac{\pi B\sin 2\eta}{4\alpha}\right\rfloor. \label{PMAX}
\end{equation}
We evaluated the eigenvalues $\lambda_{\pm}$ of $\mathcal{S}$ to $O(\alpha^{2},\lambda_{\pm}^{2})$ accuracy, as the actual R.H.S. of (\ref{quadratic}) is $O(\alpha^{2},\lambda_{\pm}^{2})$. The eigenvalues of $\mathcal{S}^{q_{\rm m}}$ are therefore accurate up to $O(q_{\rm m}\alpha^{2},q_{\rm m}\lambda_{\pm}^{2})$. Since $q_{\rm m} = \pi/\Delta\lambda = O(\lambda_{\pm}^{-1})$, the $O(q_{\rm m}\lambda_{\pm}^{2})$ terms have the same size as the $O(\lambda_{\pm}/\theta_{\rm min})$ corrections to the eigenvectors of $\mathcal{S}$. We will see in what follows that useful applications of our algorithm have $q_{\rm m}=\Theta(1/\alpha)$, in contrast to the $O(1/\alpha^{2})$ query complexity of classical search algorithms. That makes $O(q_{\rm m}\alpha^{2})$ terms $O(\alpha)$, and overall validity of the results (\ref{PMAX}) is up to $O(\alpha,\lambda/\theta_{\rm min})$.

The net query complexity of a general algorithm, which achieves $\Theta(1)$ success probability by $O(1/P_{\rm m})$ times preparation and measurement of the state $|u\rangle$, is 
\begin{equation}
Q = \frac{q_{\rm m}}{P_{\rm m}} = \frac{\pi}{4\alpha}\frac{B^{3}\sin^{2}\frac{\phi}{2}}{\sin 2\eta}\ . \label{querycomplexity}
\end{equation}
Note that quantum amplitude amplification is a faster way to take $|u\rangle$ to $|t\rangle$, by applying $\Theta(1/\sqrt{P_{\rm m}})$ times iterations of $-I_{u}I_{t}$ to it. With $I_{u} = \mathcal{S}^{q_{\rm m}}I_{s}\mathcal{S}^{-q_{\rm m}}$, the net query complexity then would be $\Theta(q_{\rm m}/\sqrt{P_{\rm m}}) = \Theta(B^{2}|\sin\frac{\phi}{2}|/\alpha)$. But $I_{u}$ requires implementation of $I_{s}$, and the construction of our general algorithm started with the hypothesis that we have only the operator $D_{s}$ available and not $I_{s}$. Without quantum amplitude amplification, we are restricted to the query complexity (\ref{querycomplexity}).

A particularly interesting case is $A = 0$, whence (\ref{etadefine}) gives $\eta = \frac{\pi}{4}$. Then
\begin{equation}
P_{\rm m} = \frac{1}{B^{2}\sin^{2}\frac{\phi}{2}}\ ,\ q_{\rm m} = \left\lfloor\frac{\pi B}{4\alpha}\right\rfloor,\ Q = \frac{\pi B^{3}}{4\alpha}\sin^{2}\frac{\phi}{2}\ . \label{PMAXA0}
\end{equation} 
With $\cos \eta = \sin \eta = \frac{1}{\sqrt{2}}$, (\ref{texpand}) and (\ref{stateexpand}) give $|\langle w|\mathcal{S}^{q}|s\rangle| = \sin(q'\Delta\lambda/2) = \sin(2q'\alpha/B)$. Thus, $A = 0$ allows us to reach the state $|w\rangle$ by iterating $\mathcal{S}$ on $|s\rangle$, $\lfloor \pi B/4\alpha \rfloor$ times. That maximizes the overlap of $|u\rangle$ with $|t\rangle$, which is not possible for $A \neq 0$. For $\phi = \pi$, we get the best results,
\begin{equation}
P_{\rm m} = |\langle t|w\rangle|^{2} = \frac{1}{B^{2}}\ ,\ q_{\rm m} = \left\lfloor\frac{\pi B}{4\alpha}\right\rfloor,\ Q= \frac{\pi B^{3}}{4\alpha}\ . \label{PMAXA0phipi}
\end{equation}
Comparing with the optimal query complexity of Grover's algorithm, $\pi/4\alpha$, we see that our algorithm is close to optimal provided $B \not\gg 1$.
 
We point out that the difference $\Delta\lambda$ between the relevant eigenvalues governs the rate of rotation from $|s\rangle$ towards $|t\rangle$. The search algorithm accomplishes its task by flipping the sign of the relative phase between $|\lambda_{+}\rangle$ and $|\lambda_{-}\rangle$ components of the state, and hence needs $q_{\rm m}=\pi/\Delta\lambda$ iterations. The query complexity of our algorithm based on the search operator $D_{s}$ is larger than $q_{\rm m}$, because the target state $|t\rangle$ is reached with probability less than one. In Section IV, we will show how to enhance this probability to $P_{\rm m}=\Theta(1)$, by controlling the operations of our algorithm with an ancilla qubit. Those manipulations improve the query complexity to $Q = \pi B/4\alpha$, closer to the optimal result of Grover's algorithm.

We also note that any practical discrete oracle should distinguish the target state sufficiently well from the non-target states. Then $\phi = \Theta(1)$, and indeed $\phi = \pi$ gives the best results. On the other hand, our analysis is valid for any value of $\phi$. 

Another special case is when $D_{s}$ is a real orthogonal operator $R_{s}$. Then its non-real eigenvalues come in complex conjugate pairs $e^{\imath\theta_{\ell_{+}}} = e^{-\imath\theta_{\ell_{-}}}$, and the corresponding eigenstates satisfy $|\ell_{+}\rangle = |\ell_{-}\rangle^{*}$. The cancellations of $\ell_{\pm}$ contributions, and $\cot\frac{\theta_{\ell}}{2} = 0$ for $\theta_{\ell} = \pi$, make $\Lambda_{1} = 0$ and $A = \cot\frac{\phi}{2}$. Then the validity condition (\ref{validitycondition2}) becomes $2(\cot\frac{|\phi|}{2}+\alpha B) \ll \theta_{\rm min}B^{2}$. Also, (\ref{etadefine}) gives
\begin{equation}
\sin 2\eta = (1+\cot^{2}2\eta)^{-1/2} = 2\alpha B/\sqrt{4\alpha^{2}B^{2}+\cot^{2}{\textstyle \frac{\phi}{2}}}\ .
\end{equation} 
Consequently,
\begin{displaymath}
P_{\rm m} = \frac{4\alpha^{2}}{4\alpha^{2}B^{2}\sin^{2}\frac{\phi}{2}+\cos^{2}\frac{\phi}{2}}\ ,
\end{displaymath}
\begin{equation}
q_{\rm m} = \frac{\pi B^{2}}{2\sqrt{4\alpha^{2}B^{2}+\cot^{2}\frac{\phi}{2}}}\ . 
\end{equation}
With $\phi = \pi$, $A = \cot\frac{\phi}{2} = 0$ and $\{P_{\rm m},q_{\rm m},Q\}$ are given by (\ref{PMAXA0phipi}). This is the case for Ambainis' generalization of Grover's algorithm, which leads to the optimal algorithm for the element distinctness problem~\cite{ambainis}. 

Lastly, consider the situation when $|A| \gg 2\alpha B$. Then (\ref{etadefine}) means $|\cot 2\eta| \gg 1$ and $\sin 2\eta \approx 1/|\cot 2\eta| = 2\alpha B/|A| \ll 1$. As a consequence, using (\ref{querycomplexity}) and (\ref{texpand}),
\begin{equation}
Q = \frac{\pi |A|B^{2}\sin^{2}\frac{\phi}{2}}{8\alpha^{2}} = \frac{\pi B}{4\alpha}\frac{|A|}{2\alpha B}\frac{1}{|\langle t|w\rangle|^{2}}\  \gg \ \frac{\pi}{4\alpha}\ . \label{Qbound}
\end{equation}
where the last inequality follows from $B\geq 1$ and $|\langle t|w\rangle|^{2} \leq 1$. Thus, for a given $B$, our algorithm is close to optimal only when $|A| \not\gg 2\alpha B$, or when $\eta$ is not too far off from $\frac{\pi}{4}$. In this situation, which includes the particular case $A=0$, the validity condition (\ref{validitycondition2}) simplifies and becomes independent of $I_{t}^{\phi}$,
\begin{equation}
|A| \not\gg 2\alpha B \ \Longrightarrow \ 2\alpha \ll \theta_{\rm min}B\ . \label{successcondition}
\end{equation}
When an estimate of $\Lambda_1$ is available, then it may be possible to adjust $\phi$ to satisfy this optimality constraint. Even for $|A| \gg 2\alpha B$, as long as $|A| \ll |\langle t|w\rangle|^{2} = B^{2}\sin^{2}\frac{\phi}{2}$, (\ref{Qbound}) gives $Q \ll \pi/8\alpha^{2}$ and our algorithm is significantly better than classical search algorithms that have query complexity $\Theta(1/\alpha^{2})$.

\subsection{Another general search operator}

Time reversal symmetry allows us to extend our analysis to the search operator $\mathcal{T} = I_{s}^{\varphi}D_{t}$, where $D_{t}$ has $|t\rangle$ as its eigenstate with eigenvalue $1$. If $D_{t}$ has an $M$-dimensional degenerate eigenspace with eigenvalue $1$, orthonormally spanned by $|t_{m}\rangle$ with $m \in \{1,2,\ldots,M\}$, then we consider the target state $|t\rangle$ to be
\begin{equation}
|t\rangle = \frac{1}{\alpha}\sum_{m}\langle t_{m}|s\rangle|t_{m}\rangle\ ,\ \ \alpha^{2} = |\langle t|s\rangle|^{2} = \sum_{m}|\langle t_{m}|s\rangle|^{2}\ . \label{multipletargetdefine}
\end{equation}
The performance of the algorithm is governed by $|\langle t|(I_{s}^{\varphi}D_{t})^{q}|s\rangle| = |\langle s|(D_{t}^{\dagger}I_{s}^{-\varphi})^{q}|t\rangle|$. That can be mapped to $|\langle t|(D_{s}I_{t}^{\varphi})^{q}|s\rangle|$ by the interchanges 
\begin{equation}
|s\rangle \leftrightarrow |t\rangle,\ \ \ \ \phi \rightarrow -\varphi,\ \ \ \ D_{s}\rightarrow D_{s}^{\dagger},\ \Longrightarrow\ \theta_{\ell} \rightarrow -\theta_{\ell}\ .
\end{equation}
With $\{|j\rangle,\theta_{j}\}$ specifying the eigenspectrum of $D_{t}$,
\begin{displaymath}
A = -(\cot\frac{\varphi}{2} + \Lambda_{1}),\ B^{2} = 1+\Lambda_{2}\ ,
\end{displaymath}
\begin{equation}
\Lambda_{p} = \sum_{j \neq t_{m}}|\langle j|s\rangle|^{2}\cot^{p}\frac{\theta_{j}}{2}\ ,\ \ \ p\in\{1,2\}\ . \label{mathcalTAB}
\end{equation}
The interchange keeps $\alpha = |\langle t|s\rangle|$ invariant, and our analysis holds as long as the condition (\ref{validitycondition2}) is satisfied by $\theta_{\rm min} = \min_{j\neq t_{m}}|\theta_{j}|$. The sign change $A \rightarrow -A$ can be accommodated by $\eta \rightarrow \frac{\pi}{2}-\eta$, and the expressions for $\{P_{\rm m},q_{\rm m},Q\}$ in (\ref{PMAX}) and (\ref{querycomplexity}) are applicable with the only change being $\phi \rightarrow -\varphi$. An interesting feature of the evolution using the search operator $\mathcal{T}$ is that the $|\lambda_{\pm}\rangle$-subspace almost completely contains the target state $|t\rangle$ but not the initial state $|s\rangle$.

\section{SPECIAL CASES}

The general quantum search algorithm presented in Section II encapsulates several previous generalizations of Grover's algorithm. In this section, we discuss some of them as special cases, providing a different perspective on their results.

\subsection{Grover's algorithm}

In Grover's algorithm, the search operator is $\mathcal{G} = -I_{s}I_{t}$. To discuss it as a special case of our general algorithm, we consider the search operator $\mathcal{G}_{\varphi,\phi} = e^{-\imath\varphi}I_{s}^{\varphi}I_{t}^{\phi}$ with selective phase rotations $\varphi$ and $\phi$ ($\mathcal{G}_{\pi,\pi} \equiv \mathcal{G}$). Then $D_{s} = e^{-\imath\varphi}I_{s}^{\varphi}$, $e^{\imath \theta_{\ell}} = e^{-\imath \varphi (1-\delta_{\ell,s})}$ and $\theta_{\rm min} = |\varphi|$. With $\theta_{s} = 0$ and $\theta_{\ell\neq s} = -\varphi$, (\ref{Omega}) gives
\begin{equation}
A = \cot\frac{\phi}{2}-\cot\frac{\varphi}{2}  = \frac{\sin\frac{\varphi - \phi}{2}}{\sin\frac{\phi}{2}\sin\frac{\varphi}{2}}\ ,\ \ B^{2} = \csc^{2}\frac{\varphi}{2}\ .
\end{equation}
For a fixed $B$, the constraint (\ref{successcondition}) demands $A \not\gg 2\alpha B$ for a nearly optimal algorithm. That requires $|\phi - \varphi| \not\gg 4\alpha\sin\frac{|\phi|}{2}$, or $|\phi - \varphi| \ll 1$ with $\alpha \ll 1$. This is the phase-matching condition~\cite{phase1,phase2}. When it is satisfied, the validity condition (\ref{validitycondition2}) of our analysis is also satisfied, because $\alpha \ll 1$ and $\theta_{\rm min}B = \varphi \csc\frac{\varphi}{2} \geq 2$. Note that $\mathcal{G}_{\varphi,\phi}$ is a special case of $\mathcal{T} = I_{s}^{\varphi}D_{t}$ as well, so our analysis also holds in case of multiple target states, if we consider $|t\rangle$ to be given by (\ref{multipletargetdefine}).

For $\phi = \varphi$, the phase-matching condition is certainly satisfied and $\eta = \pi/4$. Then (\ref{solutions}), (\ref{texpand}) and (\ref{slambdapmexpansion}) yield
\begin{eqnarray}
&\lambda_{\pm} = \pm 2\alpha\sin{\textstyle \frac{\phi}{2}}\ ,& \nonumber \\
&\langle \lambda_{\pm}|t\rangle = \langle \lambda_{\pm}|w\rangle = 1/\sqrt{2}\ ,& \nonumber \\ 
&\langle \lambda_{\pm}|s\rangle = \pm e^{-\imath \phi/2} e^{\pm \imath \alpha \sin\frac{\phi}{2}}/\sqrt{2}\ .&
\end{eqnarray} 
The evolution is thus totally confined to the $|\lambda_{\pm}\rangle$-subspace. Also, (\ref{tSqs}), (\ref{PMAX}) and (\ref{querycomplexity}) provide
\begin{displaymath}
|\langle t|\mathcal{G}_{\phi,\phi}^{q}|s\rangle| = \sin[2\alpha (q+{\textstyle \frac{1}{2}})\sin{\textstyle \frac{|\phi|}{2}}],
\end{displaymath}
\begin{equation}
P_{\rm m} = 1,\ \ Q = q_{\rm m} = \lfloor\pi \csc{\textstyle \frac{|\phi|}{2}}/4\alpha\rfloor\ . \label{specialcase2}
\end{equation}  
For $\phi = \varphi = \pi$, we have Grover's algorithm. Then $|\langle t|\mathcal{G}^{q}|s\rangle| = \sin[(2q+1)\alpha]$ and $Q = \lfloor \pi/4\alpha \rfloor$. 

\subsection{Kato's algorithm}

Let the $N$-dimensional Hilbert space be spanned by $n=\log_{2}N$ qubits, with each basis state $|j\rangle$ corresponding to an $n$-bit binary string $j \in \{0,1\}^{n}$. Then the operator $I_{s}$ cannot be written as a tensor product of single qubit operators. Its implementation needs coupling among qubits, which may be difficult to realize physically. Kato's algorithm~\cite{kato} replaces $I_{s}$ by $K_{s}$, which is a tensor product of single qubit operators. The algorithm iterates the operator $K_{s}I_{t}^{\phi}$ on $|s\rangle = H^{\otimes n}|0\rangle$, with $K_{s} = (H\mathcal{Z}_{\gamma}H)^{\otimes n}$, where
\begin{equation}
H = \frac{1}{\sqrt{2}}\left( \begin{array}{cc} 1 & 1 \\ 1 & -1\end{array} \right),\ \ \mathcal{Z}_{\gamma} = e^{-\imath \gamma}\left( \begin{array}{cc}e^{\imath \gamma} & 0 \\ 0 & e^{-\imath \gamma} \end{array} \right),
\end{equation}
and the phase $\gamma \in [-\frac{\pi}{n},\frac{\pi}{n}]$ as determined below.

The eigenvectors $|\ell\rangle$ of $K_{s}$ are $|\ell\rangle = H^{\otimes n}|j\rangle$, with the corresponding eigenphases $\theta_{\ell} = (\pi-2\gamma h_{j})_{{\rm mod}\ 2\pi} - \pi$, where $h_{j} \in \{0,\ldots,n\}$ is the Hamming weight of the binary string $j$. For the binary string of all zeroes, $h_{0} = 0$, and so $\theta_{\ell =s} =0$. Then the restriction $|\gamma| \leq \frac{\pi}{n+1}$ ensures $\theta_{\rm min} = 2|\gamma|$. The degeneracy of $\theta_{\ell}$ is $C_{n,h_{j}}$, the choose function for selecting $h_{j}$ items out of $n$. Also, $|\langle \ell|t\rangle|^{2} = 2^{-n}$ for all $\ell$. With all these values, (\ref{Omega}) gives
\begin{equation}
\Lambda_{p} = (-1)^{p}\sum_{h_{j}= 1}^{n}2^{-n}C_{n,h_{j}} \cot^{p}\gamma h_{j}\ ,\ p\in \{1,2\}.  \label{katolambdafirst}
\end{equation}
The binomial weight $2^{-n}C_{n,h_{j}}$ is maximum at $h_{j} = \frac{n}{2}$, and for large $n$ decays exponentially away from it as $\sqrt{2/\pi n}\ \exp[-2\bar{h}_{j}^{2}/n]$, with $\bar{h}_{j}= h_{j} - \frac{n}{2}$. The significant terms in the sum over $h_{j}$ therefore correspond to $\bar{h}_{j} = O(\sqrt{n}) \ll \frac{n}{2}$ for large $n$. 
To evaluate the sum in (\ref{katolambdafirst}), we use second order Taylor expansion of $\cot^{p}\gamma h_{j}$ around $\gamma n/2$,
\begin{displaymath}
(-1)^{p}\Lambda_{p} \approx  \cot^{p}\frac{\gamma n}{2} + \frac{\gamma^{2}n}{8}\left(\frac{{\rm d}^{2}\cot^{p}y}{{\rm d}y^{2}}\right)_{y=\frac{\gamma n}{2}}
\end{displaymath}
\begin{displaymath}
= \cot^{p}\frac{\gamma n}{2} + \frac{\gamma^{2}n}{4}\csc^{2}\frac{\gamma n}{2}\left[(2p-1)\cot^{p}\frac{\gamma n}{2} + \delta_{p,2}\right], 
\end{displaymath} 
where we have used the moments $\sum_{h_{j}}2^{-n}C_{n,h_{j}}\bar{h}_{j}^{r} = S_{r}$ with $S_{0} = 1$, $S_{1} = 0$ and $S_{2} = n/4$. Using the bound $\cot |\frac{\gamma n}{2}| < \csc |\frac{\gamma n}{2}| \leq |\frac{\pi}{\gamma n}|$, we get
\begin{displaymath}
-\Lambda_{1} = \cot\frac{\gamma n}{2} + O(\gamma)\csc^{2}\frac{\gamma n}{2} = \cot\frac{\gamma n[1-O(1/n)]}{2}\ ,
\end{displaymath}
\begin{equation}  
\Lambda_{2} = \left[1+O\left(1/n\right)\right]\cot^{2}\frac{\gamma n}{2}+O\left(1/n\right)\ . \label{katolambda}
\end{equation} 

Kato chose $\gamma$ such that $A = \cot\frac{\phi}{2}+\Lambda_{1} = 0$. Then (\ref{katolambda}) gives 
\begin{eqnarray}
&\gamma = (\phi/n)[1+O(1/n)],& \nonumber \\
&\Lambda_{2} = [1+O(1/n)]\cot^{2}{\textstyle \frac{\phi}{2}} + O(1/n).&
\end{eqnarray} 
Thus $B^{2} = 1+\Lambda_{2} = [1+O(1/n)]\csc^{2} \frac{\phi}{2}+O(1/n)$, and the validity condition (\ref{validitycondition2}) becomes $2\alpha \ll \theta_{\rm min}B \approx 2\gamma\csc\frac{\phi}{2} \leq 2\pi/n$ for large $n$. It is certainly satisfied because $\alpha = |\langle s|t\rangle| = 2^{-n/2}$. The performance of the algorithm follows from (\ref{PMAXA0}),
\begin{eqnarray}
&P_{\rm m} = 1-O\left(\textstyle \frac{1}{n}\right)\ ,& \nonumber \\
&q_{\rm m} = \left(\textstyle \frac{\pi 2^{n/2}}{4}\right)\left[\left\{1+O\left({\textstyle \frac{1}{n}}\right)\right\}\ \csc{\textstyle \frac{|\phi|}{2}}+O\left({\textstyle \frac{1}{n}}\right)\right].&
\end{eqnarray}
Comparing with (\ref{specialcase2}), we see that Kato's algorithm is almost as efficient as Grover's algorithm for large $n$. We also have the proof that its maximum success probability behaves as $1-O(1/n)$, which was shown only numerically by Kato. We point out that the above analysis does not hold for $\phi = \pi$, due to the restriction $|\gamma| \leq \frac{\pi}{n+1}$ that demands $|\phi| = n|\gamma|(1-O(\frac{1}{n})) \leq \pi(1-\Theta(\frac{1}{n}))$.

\subsection{Two-dimensional spatial search}

Consider a two-dimensional square lattice, whose sites encode the basis states of an $N$-dimensional Hilbert space $\mathcal{H}_{N}$. Let the sites be labeled by their coordinates as $|x,y\rangle$ for $x,y \in \{0,\ldots,\sqrt{N}-1\}$, and let $|x_{t},y_{t}\rangle$ be the label of the target state. A local operator $L$ is an operator that does not couple any pair of non-neighboring sites, i.e. $\langle x',y'|L|x,y\rangle = 0$ if $|x'-x|+|y'-y| > 1$. The two-dimensional spatial search problem is to find $|x_{t},y_{t}\rangle$, with the constraints that in one time step we can execute either an oracle query or a local operation. Such a situation can arise when the database is spread over several distinct locations, and locality of physical interactions prevents large jumps between locations. 

In Grover's algorithm, the initial state $|s\rangle$ is a uniform superposition of all lattice sites, i.e. $|s\rangle = \frac{1}{\sqrt{N}}\sum_{x,y}|x,y\rangle$. To prepare it, we start from a particular site, say $|0,0\rangle$, and then repeatedly transfer amplitudes to neighboring sites using a local operator. As we must transfer amplitude across all $\sqrt{N}$ sites in both directions of the lattice, the preparation takes $2\sqrt{N}$ local operations. Thus if $|s\rangle = \mathcal{W}|0,0\rangle$ then $\mathcal{W}$ takes $\Theta(\sqrt{N})$ time steps, and the same is true for $I_{s} = \mathcal{W}I_{0,0}\mathcal{W}^{\dagger}$. Since Grover's algorithm requires one application of $I_{s}$ per query, its total time complexity becomes $O(\sqrt{N}\times \sqrt{N}) = O(N)$, which is no better than that of classical algorithms.

To obtain quantum speedup for spatial search problems, Grover's algorithm has been modified in several ways~\cite{spatiald,spatialc,aaronson}. We look at the particular modification due to Ambainis, Kempe, and Rivosh (AKR)~\cite{spatiald}. Their algorithm uses a $4$-dim ancilla coin space $\mathcal{H}_{c}$, whose basis states $|\rightarrow\rangle, |\leftarrow\rangle, |\uparrow\rangle, |\downarrow\rangle$ represent the four possible directions of movement on a $2$-dim lattice. Let $|u_{c}\rangle$ be the uniform superposition of these coin basis states. AKR's algorithm is implemented in the $4N$-dim Hilbert space $\mathcal{H}_{c}\otimes \mathcal{H}_{N}$. It drives the initial state $|s'\rangle \equiv |u_{c}\rangle|s\rangle$ towards the effective target state $|t'\rangle \equiv |u_{c}\rangle|x_{t},y_{t}\rangle$ by iterating the operator $L_{s'}I_{t'}$. Here $I_{t'} = \mathbbm{1}_{4N} - 2|t'\rangle\langle t'|$ is implemented using an oracle query, while $L_{s'} = \mathcal{L}\bar{I}_{u_{c}}$ is a local operation with $\bar{I}_{u_{c}} = 2|u_{c}\rangle\langle u_{c}| - \mathbbm{1}_{4}$ and $\mathcal{L}$ defined by
\begin{eqnarray}
&|\rightarrow\rangle \otimes |x,y\rangle  \stackrel{\mathcal{L}}{\longrightarrow}  |\leftarrow\rangle \otimes |x+1,y\rangle, & \nonumber \\
&|\leftarrow\rangle \otimes |x,y\rangle  \stackrel{\mathcal{L}}{\longrightarrow}  |\rightarrow\rangle \otimes |x-1,y\rangle, & \nonumber \\
&|\uparrow\rangle \otimes |x,y\rangle  \stackrel{\mathcal{L}}{\longrightarrow}  |\downarrow\rangle \otimes |x,y+1\rangle,& \nonumber \\  
&|\downarrow\rangle \otimes |x,y\rangle  \stackrel{\mathcal{L}}{\longrightarrow} |\uparrow\rangle \otimes |x,y-1\rangle.& \label{movingstep}
\end{eqnarray}
AKR found the eigenstates of $L_{s'}$ to be 
\begin{equation}
|\ell_{a,b}\rangle = |v_{a,b}\rangle |\mathcal{F}_{a}\rangle |\mathcal{F}_{b}\rangle,\ \ a,b\in \{0,1,\ldots,\sqrt{N}-1\},
\end{equation}
where $|\mathcal{F}_{a}\rangle = N^{-\frac{1}{4}}\sum_{x}e^{\frac{\imath 2\pi a\cdot x}{\sqrt{N}}}|x\rangle$ and $|\mathcal{F}_{b}\rangle = N^{-\frac{1}{4}}\sum_{y}e^{\frac{\imath 2\pi b\cdot y}{\sqrt{N}}}|y\rangle$ form the Fourier basis. For each $(a,b)$, there are four eigenvalues $1$, $-1$, and $e^{\pm \imath\theta_{a,b}}$ with 
\begin{equation}
2\cos\theta_{a,b}=\cos(2\pi a/\sqrt{N})+\cos(2\pi b/\sqrt{N}), \label{spatialeigenangle}
\end{equation} 
corresponding to four orthogonal vectors $|v_{a,b}^{1}\rangle$, $|v_{a,b}^{-1}\rangle$ and $|v_{a,b}^{\pm}\rangle$ of the coin space. 

AKR have shown that $L_{s'}I_{t'}$ preserves the subspace $\mathcal{H}_{0}$ spanned by the eigenstates $|s'\rangle = |u_{c}\rangle|\mathcal{F}_{0}\rangle|\mathcal{F}_{0}\rangle$ and $|\ell_{a,b}^{\pm}\rangle =|v_{a,b}^{\pm}\rangle|\mathcal{F}_{a}\rangle|\mathcal{F}_{b}\rangle$ with $(a,b)\neq (0,0)$. Furthermore, the states $|v_{a,b}^{\pm}\rangle$ are such that $|\langle \ell_{a,b}^{\pm}|t'\rangle| = 1/\sqrt{2N}$. 
Within $\mathcal{H}_{0}$, $|s'\rangle$ is a unique eigenstate with eigenvalue $1$, and $L_{s'}$ is a real orthogonal operator with $A = 0$ as described in Section II.D. To compute $B$, we put $|\langle \ell_{a,b}^{\pm}|t'\rangle| = 1/\sqrt{2N}$ and $1+\cot^{2}x = \csc^{2}x = 2/(1-\cos 2x)$ in (\ref{Omega}), to get
\begin{eqnarray}
B^{2} &=& \frac{2}{N}\sum_{(a,b)\neq (0,0)}\frac{1}{1-\cos\theta_{a,b}} \nonumber \\
      &=& \frac{4}{N}\sum_{(a,b)\neq (0,0)} \left[2-\cos\frac{2\pi a}{\sqrt{N}}-\cos\frac{2\pi b}{\sqrt{N}}\right]^{-1}\ . 
\end{eqnarray} 
Using $1-\cos x = 2\sin^{2}\frac{x}{2}$, and $x \geq \sin x \geq \frac{2}{\pi}x$ for $x \in [0,\frac{\pi}{2}]$,
\begin{eqnarray}
B^{2} &=& \frac{2}{N}\sum_{(a,b)\neq (0,0)}\left[\sin^{2}\frac{\pi a}{\sqrt{N}}+\sin^{2}\frac{\pi b}{\sqrt{N}}\right]^{-1} \nonumber \\
      &=& \Theta\left(\sum_{(a,b)\neq(0,0)}\frac{1}{a^{2}+b^{2}}\right) = \Theta(\ln N). \label{twodimspatialB}
\end{eqnarray}
Here the last equality follows from $\sum_{b}(a^{2}+b^{2})^{-1} = \Theta(1/a)(1-\delta_{a,0})+\Theta(1)\delta_{a,0}$, and $\sum_{a = 1}^{\sqrt{N}-1}\Theta(1/a) = \Theta(\ln N)$. Also, (\ref{spatialeigenangle}) implies that $\theta_{\rm min} = \min_{(a,b)\neq(0,0)}\theta_{a,b} = \sqrt{2}\pi/\sqrt{N}$. The condition (\ref{validitycondition2}) for the validity of our analysis is satisfied for large $N$, because $A = 0$ and $2\alpha = 2/\sqrt{N} \ll \theta_{\rm min}B = \sqrt{\Theta(\ln N)/N}$. Thereafter, with $\alpha = 1/\sqrt{N}$ and $|w'\rangle = (L_{s'}I_{t'})^{q_{\rm m}}|s'\rangle$, (\ref{PMAXA0phipi}) gives
\begin{equation}
P_{\rm m} = \langle t'|w'\rangle^{2} = \Theta(1/\ln N)\ ,\ q_{\rm m} = \Theta(\sqrt{N\ln N}). 
\end{equation}
Note that our analysis gives a more accurate estimate of $P_{\rm m}$ compared to AKR's estimate, $P_{\rm m} = \Omega(1/\ln N)$. 
 
Subsequently, AKR's algorithm uses quantum amplitude amplification algorithm to reach $|t'\rangle$ by applying $\Theta(1/|\langle t'|w'\rangle|) = \Theta(\sqrt{\ln N})$ iterations of $-I_{w'}I_{t'}$ on $|w'\rangle$, with $I_{w'} = (L_{s'}I_{t'})^{q_{\rm m}}I_{s'}(L_{s'}^{\dagger}I_{t'}^{\dagger})^{q_{\rm m}}$. Now $I_{s'}$ can be implemented in $\Theta(\sqrt{N})$ time steps just like $I_{s}$, so $I_{w'}$ takes $\Theta(q_{\rm m})+\Theta (\sqrt{N}) = \Theta (\sqrt{N\ln N})$ time steps. The net query complexity of the algorithm is, therefore, $\Theta(\sqrt{N\ln N} \times \sqrt{\ln N}) = \Theta (\sqrt{N}\ln N)$. 

It was an open question whether or not the $\Theta (\sqrt{N}\ln N)$ algorithm for $2$-dim spatial search can be improved further. In the next Section, we give a positive answer to this question by presenting a $\Theta (\sqrt{N\ln N})$ algorithm, which is $\Theta (\sqrt{\ln N})$ times faster than AKR's algorithm.

\section{CONTROLLED QUANTUM SEARCH}

In Section II, we analysed the performance of the general search operator $\mathcal{S} = D_{s}I_{t}^{\phi}$. In this Section, we show that just by controlling the operators $\{D_{s},I_{t}^{\phi}\}$ through an ancilla qubit, without modifying them at all, we can manipulate the moments $\{\Lambda_{1},\Lambda_{2}\}$ and hence the coefficients $\{A,B\}$. The results (\ref{PMAX}) and (\ref{querycomplexity}) imply that the best performance of the general quantum search algorithm is achieved by $\phi = \pi$, $A = 0$ and $B \geq 1$ as small as possible. The aim of our manipulations is to achieve that situation.

\subsection{Manipulation of the first moment}

To control the search process, we attach an ancilla qubit with the Hilbert space $\mathcal{H}_{2}$ to the search space $\mathcal{H}_{N}$, and work in the joint space $\mathcal{H}' = \mathcal{H}_{2} \otimes \mathcal{H}_{N}$. The logic circuit of Algorithm $1$ that effectively makes $A$ vanish is shown in Fig.~1, where the dashed box represents the search operator $\mathcal{S}' = D_{s}'I_{t'}$ with
\begin{eqnarray}
&D_{s}' = (c_{1}D_{s}^{\dagger})(c_{0}D_{s}),& \nonumber \\ 
&c_{0}D_{s} = |0\rangle\langle 0| \otimes D_{s} + |1\rangle\langle 1| \otimes \mathbbm{1}_{N}\ ,&\nonumber \\ 
&c_{1}D_{s}^{\dagger} = |0\rangle\langle 0| \otimes \mathbbm{1}_{N} + |1\rangle\langle 1| \otimes D_{s}^{\dagger},& \label{controlDsprime}\nonumber \\
&I_{t'} = \mathbbm{1}_{2N} - 2|t'\rangle\langle t'|,\ \ |t'\rangle = |+\rangle|t\rangle,& \label{tprimedefine}
\end{eqnarray}
where $|+\rangle = {\textstyle \frac{1}{\sqrt{2}}}[|0\rangle +|1\rangle]$.
Here $c_{0}D_{s}$ and $c_{1}D_{s}^{\dagger}$ control the operation of $D_{s}$ or $D_{s}^{\dagger}$ on $\mathcal{H}_{N}$ through the ancilla qubit. The eigenstates of $D_{s}'$ are $|\ell_{0}'\rangle =|0\rangle|\ell\rangle$ and $|\ell_{1}'\rangle = |1\rangle|\ell\rangle$, with $\theta_{\ell}$ and $-\theta_{\ell}$ as the corresponding eigenphases. Thus
\begin{equation}
\theta_{\ell_{0}}' = \theta_{\ell}\ ,\ \theta_{\ell_{1}}' = -\theta_{\ell}\ ,\ \ \langle \ell_{0}'|t'\rangle = \langle\ell_{1}'|t'\rangle = {\textstyle \frac{1}{\sqrt{2}}}\langle \ell|t\rangle. \label{firstmanipulationvalues}
\end{equation}
$D_{s}'$ has two mutually orthogonal eigenstates, $|s_{0}'\rangle = |0\rangle|s\rangle$ and $|s_{1}'\rangle = |1\rangle|s\rangle$, with eigenvalue $1$. In accordance with (\ref{initialstatedefine}), we choose the initial state $|s'\rangle$ to be 
\begin{eqnarray}
&|s'\rangle = {\textstyle \frac{1}{\alpha'}}\ [\langle s_{0}'|t'\rangle |s_{0}'\rangle  + \langle s_{1}'|t'\rangle |s_{1}'\rangle] = |+\rangle|s\rangle,& \nonumber \\ 
&\alpha'  = \sqrt{|\langle s_{0}'|t'\rangle|^{2}+|\langle s_{1}'|t'\rangle|^{2}} = |\langle s|t\rangle|= \alpha.&
\end{eqnarray}
We also have $\theta_{\rm min}' = \min_{\ell' \neq s_{m}'}\theta_{\ell'} = \theta_{\rm min}$. Now, putting (\ref{firstmanipulationvalues}) in (\ref{Omega}), we get
\begin{displaymath}
\Lambda_{p}'= \sum_{\ell_{0}' \neq s_{0}'}|\langle \ell_{0}'|t'\rangle|^{2}\cot^{p}\frac{\theta_{\ell_{0}}'}{2} + \sum_{\ell_{1}' \neq s_{1}'}|\langle \ell_{1}'|t'\rangle|^{2}\cot^{p}\frac{\theta_{\ell_{1}}'}{2} 
\end{displaymath}
\begin{equation}
  \Longrightarrow \ \Lambda_{p}' \ = \ \delta_{p,2}\Lambda_{p}\ .
\end{equation}
Explicitly, $A' = \cot\frac{\phi}{2}+\Lambda_{1}' = 0$ for $\phi = \pi$, and $B' = \sqrt{1+\Lambda_{2}'} = B$. Since $\{\theta_{\rm min}',\alpha',B'\} = \{\theta_{\rm min},\alpha,B\}$, whenever $D_{s}$ satisfies the condition $\theta_{\rm min}B \gg 2\alpha$, its equivalent condition $\theta_{\rm min}'B' \gg 2\alpha'$ is satisfied by $D_{s}'$. The results of Section II hold, therefore, for analysing the iteration of $D_{s}'I_{t'}$ on $|s'\rangle$ to reach the effective target state $|t'\rangle$, whose projection on $\mathcal{H}_{N}$ is the actual target state $|t\rangle$. As $A' =  0$, the corresponding $\{P_{\rm m}',q_{\rm m}', Q'\}$ are given by (\ref{PMAXA0}), i.e.
\begin{equation}
P_{\rm m}' = B^{-2}\ ,\ \ q_{\rm m}' = \lfloor \pi B/4\alpha\rfloor,\ \ Q' = \pi B^{3}/4\alpha\ . \label{firstmanipulationresult}
\end{equation}
The importance of this manipulation is that $A'$ is always zero irrespective of the value of $A$, so that we get the optimal algorithm for fixed $\{\theta_{\rm min}, \alpha, B\}$. 

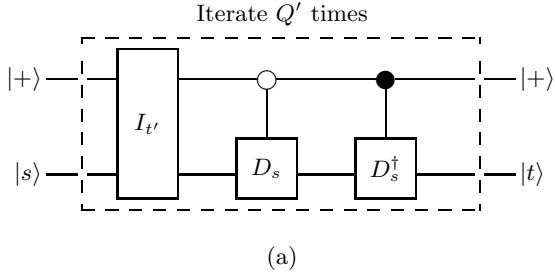
\begin{figure}[t!]
\setlength{\unitlength}{0.9pt}
\begin{picture}(500,100)(50,-10)
\put(83,25){\makebox(0,0)[r]{$|s\rangle$}}
\put(83,65){\makebox(0,0)[r]{$|+\rangle$}}
\put(284,25){\makebox(0,0)[l]{$|t\rangle$}}
\put(284,65){\makebox(0,0)[l]{$|+\rangle$}}
\put(85,25){\line(1,0){13}}
\put(102,25){\line(1,0){13}}
\put(140,25){\line(1,0){25}}
\put(190,25){\line(1,0){25}}
\put(240,25){\line(1,0){25}}
\put(269,25){\line(1,0){13}}
\put(85,65){\line(1,0){13}}
\put(102,65){\line(1,0){13}}
\put(140,65){\line(1,0){33.5}}
\put(181.5,65){\line(1,0){42}}
\put(231.5,65){\line(1,0){33.5}}
\put(269,65){\line(1,0){13}}
\put(165,15){\framebox(25,25){$D_{s}$}}
\put(215,15){\framebox(25,25){$D_{s}^{\dagger}$}}
\put(115,15){\framebox(25,62.5){$I_{t'}$}}
\put(177.5,61){\line(0,-1){21}}
\put(227.5,61){\line(0,-1){21}}
\put(177.5,65){\circle{8}}
\put(227.5,65){\circle*{8}}
\put(100,10){\dashbox{5}(167,72.5){}}
\put(184,87.5){\makebox(0,0)[b]{{Iterate $Q'$ times}}}
\put(184,-15){\makebox(0,0)[b]{{(a)}}}
\end{picture}
\caption{\small Logic circuit diagram for Algorithm 1.}
\end{figure}

\subsection{Manipulation of the second moment}

The query complexity $Q'$ can be further improved from $\pi B^{3}/4\alpha$ to $\pi B/4\alpha$ by manipulating the second moment $\Lambda_{2}$. Without loss of generality, we analyse only the $\Lambda_{1} = 0$ case, obtained if necessary by addition of an ancilla qubit as in Algorithm 1. The logic circuit of Algorithm $2$ that effectively reduces $B$ is shown in Fig.~2, where the dashed box represents the search operator $\mathcal{S}'' = D_{s}''I_{t''}$ in the joint space $\mathcal{H}'' = \mathcal{H}_{2} \otimes \mathcal{H}_{N}$, with
\begin{eqnarray}
&D_{s}'' = (Z \otimes \mathbbm{1}_{N})(c_{0}D_{s}),&\ \nonumber \\
&|t''\rangle = |\zeta\rangle|t\rangle, \ |\zeta\rangle = \sin \zeta|0\rangle +\cos\zeta|1\rangle.&
\end{eqnarray} 
Here $Z = |0\rangle\langle 0| -|1\rangle \langle 1|$ is the Pauli $Z$ operator. The eigenstates of $D_{s}''$ are $|\ell_{0}''\rangle =|0\rangle|\ell\rangle$ and $|\ell_{1}''\rangle = |1\rangle|\ell\rangle$, with $\theta_{\ell}$ and $\pi$ as the corresponding eigenphases. Thus
\begin{eqnarray}
&\theta_{\ell_{0}}'' = \theta_{\ell}\ ,\ \theta_{\ell_{1}}'' = \pi\ ,& \nonumber \\
&\langle \ell_{0}''|t''\rangle = \langle \ell|t\rangle\sin\zeta\ ,\ \langle \ell_{1}''|t''\rangle = \langle \ell|t\rangle\cos\zeta\ .& \label{secondmanipulationvalues}
\end{eqnarray}

Here, $D_{s}''$ has a unique eigenstate $|s_{0}''\rangle = |0\rangle|s\rangle$ with eigenvalue $1$, and we choose it as the initial state $|s''\rangle$. So $\alpha'' = \langle s''|t''\rangle = \alpha \sin \zeta$ and $\theta_{\rm min}'' = \min_{\ell''\neq s_{0}''}|\theta_{\ell}''| = \theta_{\rm min}$. Now, putting (\ref{secondmanipulationvalues}) in (\ref{Omega}), we get
\begin{displaymath}
\Lambda_{p}'' = \sum_{\ell_{0}'' \neq s_{0}''}|\langle \ell_{0}''|t''\rangle|^{2}\cot^{p}\frac{\theta_{\ell_{0}}''}{2} + \sum_{\ell_{1}'' \neq s_{1}''}|\langle \ell_{1}''|t''\rangle|^{2}\cot^{p}\frac{\theta_{\ell_{1}}''}{2}
\end{displaymath}
\begin{equation}
 \Longrightarrow\ \Lambda_{p}''\ =\ \Lambda_{p}\sin^{2}\zeta\ .
\end{equation}
With $\phi = \pi$ and $\Lambda_{1} =0$, we have $A'' = \Lambda_{1}'' = 0$. Furthermore,
\begin{eqnarray}
(B'')^{2} &=& 1+\Lambda_{2}'' = 1+\Lambda_{2}\sin^{2}\zeta \nonumber \\
          &=& \cos^{2}\zeta + B^{2}\sin^{2}\zeta\ \ \stackrel{\zeta \ll 1}{\approx}\ \ 1+B^{2}\zeta^{2}. \nonumber
\end{eqnarray}
For $B \gg 1$, we choose $\zeta \leq B^{-1} \ll 1$, to obtain $1 \leq (B'')^{2} \leq 2$. Also $\alpha'' \approx \alpha \zeta \leq \alpha/B$ means that $2\alpha/B \geq 2\alpha'' \geq 2\alpha''/B''$, and so whenever $D_{s}$ satisfies $\theta_{\rm min}B\gg 2\alpha$, $D_{s}''$ satisfies $\theta_{\rm min}''B'' \gg 2\alpha''$. The analysis of Section II therefore holds for $D_{s}''$, and (\ref{firstmanipulationresult}) reduces to
\begin{displaymath}
P_{\rm m}'' = \frac{1}{(B'')^{2}} \in \left[\frac{1}{2},1\right],\ \ q_{\rm m}''= \left\lfloor\frac{\pi B''}{4\alpha \zeta}\right\rfloor,
\end{displaymath}
\begin{equation}
Q''_{\zeta} = \frac{\pi}{4\alpha}\frac{(1+B^{2}\zeta^{2})^{3/2}}{\zeta}\ .
\end{equation}
Minimizing $Q_{\zeta}''$ over $\zeta$, the optimal choice is $\zeta =  1/\sqrt{2}B$, yielding
\begin{equation}
Q''_{\rm min} = Q''_{\zeta = 1/\sqrt{2}B} = (3\sqrt{3}/2)(\pi B/4\alpha)\ .
\end{equation}
Thus we have an $O(B/\alpha)$ search algorithm; the $O(B^{2})$ speedup is useful only when $B \gg 1$ that we assumed. Note that $D_{s}^{\dagger}$ is not needed here unlike in Algorithm 1.

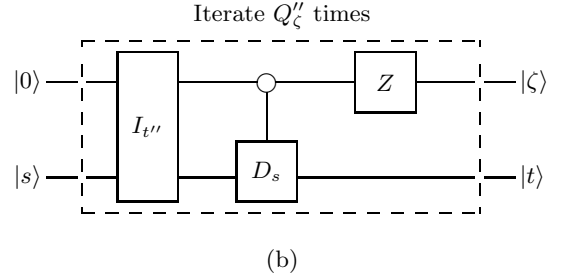
\begin{figure}[t!]
\setlength{\unitlength}{0.9pt}
\begin{picture}(500,100)(300,-10)
\put(333,25){\makebox(0,0)[r]{$|s\rangle$}}
\put(333,65){\makebox(0,0)[r]{$|0\rangle$}}
\put(534,25){\makebox(0,0)[l]{$|t\rangle$}}
\put(534,65){\makebox(0,0)[l]{$|\zeta\rangle$}}
\put(335,25){\line(1,0){13}}
\put(352,25){\line(1,0){13}}
\put(390,25){\line(1,0){25}}
\put(440,25){\line(1,0){25}}
\put(490,25){\line(1,0){25}}
\put(519,25){\line(1,0){13}}
\put(465,25){\line(1,0){25}}
\put(335,65){\line(1,0){13}}
\put(352,65){\line(1,0){13}}
\put(390,65){\line(1,0){33.5}}
\put(431.5,65){\line(1,0){33.5}}
\put(490,65){\line(1,0){25}}
\put(519,65){\line(1,0){13}}
\put(415,15){\framebox(25,25){$D_{s}$}}
\put(465,52.5){\framebox(25,25){$Z$}}
\put(365,15){\framebox(25,62.5){$I_{t''}$}}
\put(427.5,61){\line(0,-1){21}}
\put(427.5,65){\circle{8}}
\put(350,10){\dashbox{5}(167,72.5){}}
\put(434,87.5){\makebox(0,0)[b]{{Iterate $Q''_{\zeta}$ times}}}
\put(434,-15){\makebox(0,0)[b]{{(b)}}}

\end{picture}
\caption{\small Logic circuit diagram for Algorithm 2.}
\end{figure}
\subsection{Applications}

\subsubsection{Faster algorithm for 2-dim spatial search}

The manipulations of Sections IV.A and IV.B find an important application in two-dimensional spatial search, discussed in Section III.E. The spatial search is restricted by nearest neighbour movement, and in $d$ dimensions $\Omega(N^{1/d})$ steps are needed to cover the whole database. Grover's optimal search algorithm allows movement from any site to any other site in just one step, and can be effectively considered the $d \rightarrow \infty$ limit of spatial search. These considerations imply that the spatial search problem has complexity $\Omega(\max(N^{1/d},\sqrt{N}))$.

For the $d=1$ case, quantum spatial search has the same complexity as classical search. For $d>2$, one can achieve $O(\sqrt{N})$ complexity with quantum spatial search, comparable to Grover's algorithm. The critical $d=2$ case is special, with the conflict between different dynamical features producing $\ln N$ factors. We have seen that $B = \Theta(\sqrt{\ln N})$ for $d=2$, as a result of the divergence in (\ref{twodimspatialB}). (Note that $B = \Theta(1)$ for $d>2$.) With $\alpha = 1/\sqrt{N}$, the manipulations described above give an $O(B/\alpha) = O(\sqrt{N\ln N})$ algorithm, which is $O(\sqrt{\ln N})$ times faster than AKR's algorithm. 

\subsubsection{Finding eigenstate for a given eigenvalue}

Consider the problem of finding an eigenstate $|s\rangle$ of a given operator $D$ with known eigenvalue $e^{\imath \theta_{s}}$. We first construct the operator $D_{s} = e^{-\imath \theta_{s}}D$, which has eigenvalue $1$ for $|s\rangle$. The search algorithm then can take us to the state $|t\rangle$ using $D_s$, $D_s^{\dagger}$ and $I_{t}$. Reversing the algorithm, we can start with a known state $|t\rangle$, and evolve to $|s\rangle$ using $O(B/\alpha)$ applications of $\{I_{t},D,D^{\dagger}\}$ (we can use Algorithms 1 and 2 if necessary). The required property is that the eigenphases $\theta_{\ell \neq s}$ of $D$ should be well-separated from $\theta_{s}$ according to the condition (\ref{validitycondition2}), i.e. $\theta_{\rm min} = \min_{\ell \neq s}(\theta_{\ell}-\theta_{s}) \gg 2\alpha/B$. 

\section{CONCLUSION}

We have analysed a general framework of the quantum
search algorithms, where one of the selective transformations
gets replaced by a more general unitary transformation.
We have derived the conditions for a successul quantum
search and calculated the number of iterations required
by the algorithm. We have discussed several quantum
search algorithms as special cases of our general framework.
There are other search algorithms also, not discussed
here, which can be considered as special cases.
For example, the quantum random walk search algorithm
presented by Shenvi, Kempe and Whaley~\cite{shenvi} is equivalent
to spatial search in $d = \log_{2} N$ dimensions.

We have shown that the search operators can be controlled through an ancilla qubit to get faster quantum search algorithms. These algorithms
may find interesting applications. For example, a faster quantum walk algorithm for the two-dimensional spatial search can be obtained using similar techniques~\cite{faster}. Some other possibilities
are under investigation. 

We point out that the general framework presented
here applies only to iterative search algorithms and does
not apply to recursive quantum search algorithms (see,
for example, section III of ~\cite{tulsiselective}]). Our analysis provides
insights in to the nature of iterative quantum search algorithms,
where the performance depends completely on
the eigenspectrum of the search operator. That is unlike
the recursive case, where the performance depends upon
the amplification factor provided by the search operator.
We believe that our general framework will serve as
an important tool in designing future iterative quantum
search algorithms.

\textbf{Acknowledgements}: I thank Prof. Apoorva Patel
for going through the manuscript and for useful comments
and discussions.


\begin{thebibliography}{99}
\bibitem{grover} L.K. Grover, Phys. Rev. Lett. \textbf{79}, 325 (1997). 
\bibitem{qaa1} L.K. Grover, Phys. Rev. Lett. \textbf{80}, 4329 (1998).
\bibitem{qaa2} G. Brassard, P. Hoyer, M. Mosca, and A. Tapp, Contemporary Mathematics (American Mathematical Society, Providence), \textbf{305}, 53 (2002) [arXiv.org:quant-ph/0005055].
\bibitem{optimal}C. Bennett, E. Bernstein, G. Brassard, and U. Vazirani, SIAM J. Computing {\bf 26}, 1510 (1997) [arXiv.org:quant-ph/9701001].
\bibitem{phase1} G. L. Long, Y. S. Li, W. L. Zhang, and L. Niu, Phys. Lett. A {\bf 262}, 27 (1999).
\bibitem{phase2} P. Hoyer, Phys. Rev. A {\bf 62}, 052304 (2000).
\bibitem{kato} G. Kato, Phys. Rev. A \textbf{72}, 032319 (2005).
\bibitem{ambainis} A. Ambainis, SIAM J. Computing, {\bf 37}, 210 (2007) [arXiv.org:quant-ph/0311001].
\bibitem{spatiald} A. Ambainis, J. Kempe, and A. Rivosh, Proc. 16th ACM-SIAM SODA, p.~1099 (2005) [arXiv.org:quant-ph/0402107].
\bibitem{spatialc} A. Childs and J. Goldstone, Phys. Rev. A {\bf 70}, 042312 (2004).
\bibitem{aaronson} S. Aaronson and A. Ambainis, Proc. 44th IEEE Symposium on Foundations of Computer Science (IEEE, Los Alamitos, 2003), p. 200 [arXiv.org:quant-ph/0303041].
\bibitem{faster} A. Tulsi, Phys. Rev. A {\bf 78}, 012310 (2008).
\bibitem{shenvi} N. Shenvi, J. Kempe, and K. B. Whaley, Phys. Rev. A \textbf{67}, 052307 (2003).
\bibitem{tulsiselective} A. Tulsi, Phys. Rev. A {\bf 78}, 022332 (2008).
\end{thebibliography}
\end{document}